\documentclass{article}

\usepackage{PRIMEarxiv} 

\usepackage[utf8]{inputenc} 
\usepackage[T1]{fontenc}    
\usepackage{hyperref}
\usepackage{url}            
\usepackage{booktabs}       
\usepackage{bm}
\usepackage{amsfonts}       
\usepackage{amsmath}
\usepackage{amssymb} 
\usepackage{nicefrac}       
\usepackage{microtype}
\usepackage{cleveref}
\usepackage{xcolor}
\usepackage{algpseudocode}
\usepackage{algorithm}
\usepackage{color,soul}
\usepackage{caption}
\usepackage{lipsum}
\usepackage{fancyhdr}       
\usepackage{graphicx}       
\graphicspath{{media/}}     
\usepackage{enumitem}
\usepackage{comment}
\makeatletter
\newcommand{\printfnsymbol}[1]{%
  \textsuperscript{\@fnsymbol{#1}}%
}
\makeatother
\vspace{-3.5cm}
\title{TOMAS: Topology Optimization of Multiscale Fluid Devices using Variational Autoencoders and Super-Shapes}

\author{
  Rahul Kumar Padhy \\
  Department of Mechanical Engineering \\
  University of Wisconsin-Madison \\
  Madison, WI, USA \\
  \texttt{rkpadhy@wisc.edu} \\
  \And
  Krishnan Suresh \\
  Department of Mechanical Engineering \\
  University of Wisconsin-Madison \\
  Madison, WI, USA \\
  \texttt{ksuresh@wisc.edu} \\
    \And
  Aaditya Chandrasekhar \\
 Advanced Photon Source \\
 Argonne National Laboratory \\
 Lemont, IL, USA \\
\texttt{cs.aaditya@gmail.com} \\
}

\begin{document}
\maketitle

\vspace{-1.0cm}
\begin{abstract}

In this paper, we present a framework for multiscale topology optimization of fluid-flow devices. The objective is to minimize dissipated power, subject to a desired contact-area. The proposed strategy is to design optimal microstructures in individual finite element cells, while simultaneously optimizing the overall fluid flow. In particular, parameterized super-shape microstructures are chosen here to represent microstructures since they exhibit a wide range of permeability and contact area. To avoid repeated homogenization, a finite set of these super-shapes are analyzed \emph{a priori}, and a variational autoencoder (VAE) is trained on  their fluid constitutive properties (permeability), contact area and shape parameters. The resulting differentiable latent space is integrated with a coordinate neural network to carry out a global multi-scale fluid flow optimization.  The latent space enables the use of new microstructures that were not present in the original data-set. The proposed method is illustrated using numerous examples in 2D.

\begin{figure}[H]
 	\begin{center}
		\includegraphics[scale=0.35,trim={0 0 0 0},clip]{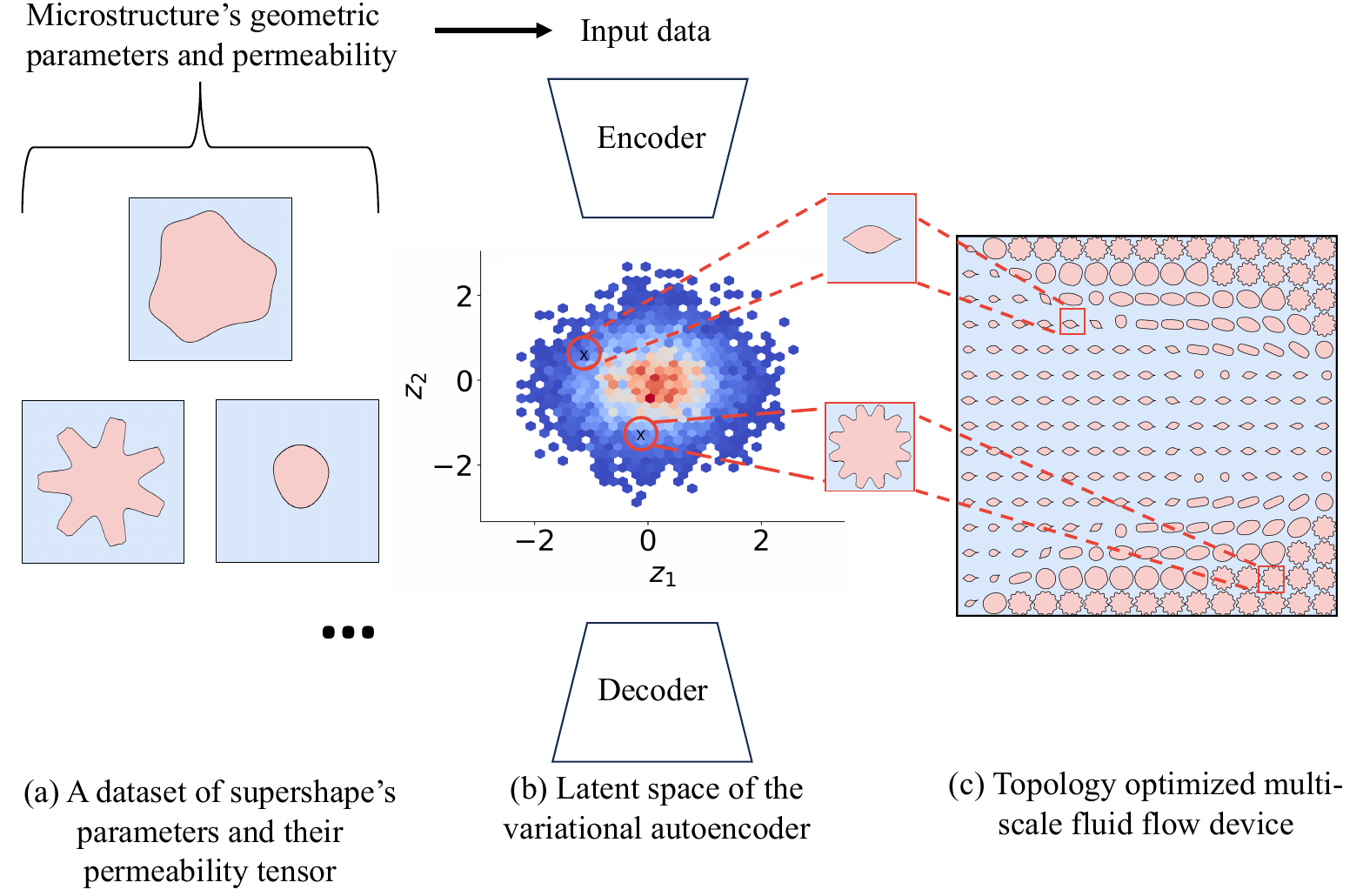}
 		\caption*{Overview of the proposed method: A dataset of shape parameters of super-shapes, contact areas, and homogenized permeability tensors is used to train a variational autoencoder (VAE). The resulting latent space is then used for global multiscale optimization of fluid flow.}
 		\label{fig:abstract}
	\end{center}
 \end{figure}
\end{abstract}


\section{Introduction}
\label{sec:intro}
\begin{figure}[H]
 	\begin{center}
	\includegraphics[scale=0.4,trim={0 0 0 0} ]{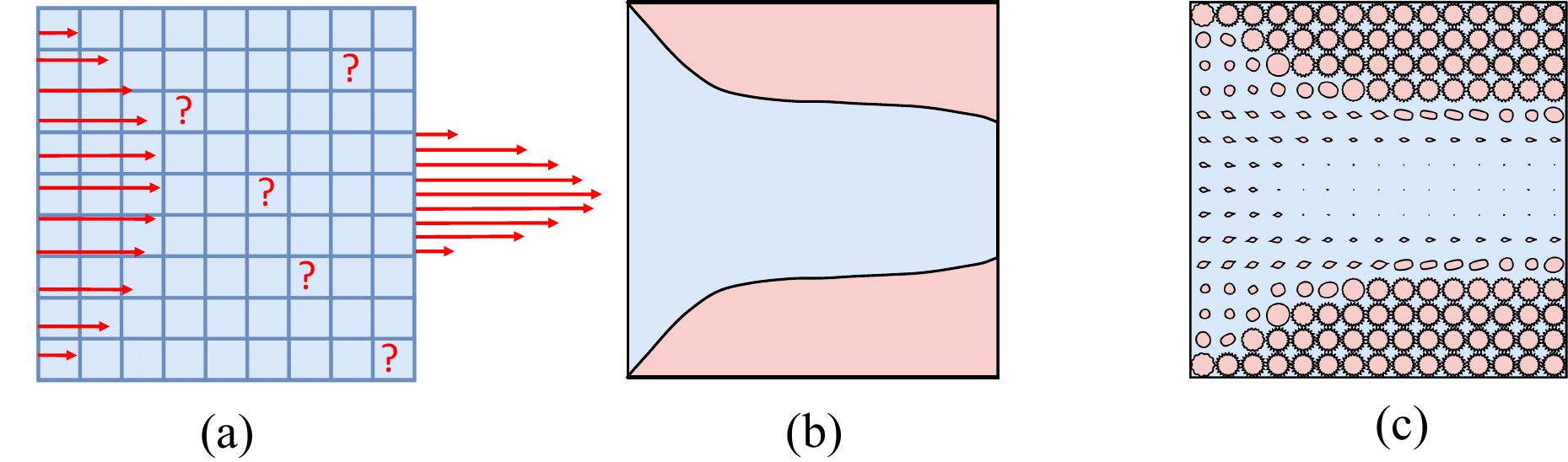}	\caption{(a) Fluid-flow problem. (b) Single-scale design. (c) Multi-scale design. }
	\label{fig:Fluid_MTO}
	\end{center}
 \end{figure}

In fluid-flow based topology optimization, the typical objective is to determine the path of least resistance, i.e., least dissipation, within a design domain; see \cref{fig:Fluid_MTO}(a). When no other constraint is imposed, the path of least resistance is a single connected path \cite{alexandersen2020review} as illustrated in \cref{fig:Fluid_MTO}(b). However, when additional constraints are introduced, the optimal flow-path is typically more complex, and not necessarily a single connected path. One such constraint is the desired fluid-solid contact area, which plays a crucial role in various applications such as bio-sensors for detecting tumor cells \cite{nagrath2007isolation}, microfluidic devices for cell sorting \cite{fan1999dynamic, hayes2001flow, jiang2000mrna, liu2007micropillar, choi2002integrated}, micro-channel heat sinks \cite{zhu2016prediction, guo2013multiphysics, moran2004microsystem}, and other microfluidic devices involving heat transfer and mass transportation/mixing mechanisms \cite{bixler2012bioinspired,bixler2013fluid}. In these applications, a minimum fluid-solid contact area is critical for achieving desired performance and functionality. For instance, in bio-sensors for detecting tumor cells, increased contact area between the fluid and the sensor surface enhances the sensitivity and accuracy of the detection process. Similarly, in microfluidic devices for cell sorting, the efficiency of cell capture and separation depends on the contact area between the cells and the solid surfaces. One approach for enhancing the contact area is to employ arrays of micro-pillars, as suggested in  \cite{huang2018review, li2014high}, but this can result in a substantial increase in dissipated power \cite{lauder2016structure, bocanegra2016holographic}.  A more powerful approach is to use multi-scale structures, illustrated in \cref{fig:Fluid_MTO}(c), and the main focus of this paper.

\subsection{Single-Scale Fluid Topology Optimization}
\label{sec: fluid_TO}
The field of fluid flow TO was initiated by the seminal work of Borrvall and Petersson \cite{borrvall2003topology}. In their pioneering research, they presented an optimal flow layout that minimizes pressure drop by employing the Stokes equation along with the Brinkman-Darcy law equations under low Reynolds number conditions. Gersborg-Hansen et al. \cite{gersborg2005topology} continued this  work by presenting applications with low Reynolds numbers for microfluidic problems and micro-electro-mechanical devices. Guest and Prévost \cite{guest2006topology} introduced the method of Darcy–Stokes finite elements to optimize creeping fluid flows, producing 0–1 (void-solid) topologies without artificial material regions.  

A novel density-based approach for topology optimization of Stokes flow was proposed in \cite{haubner2023novel} which addresses convergence issues using fractional-order Sobolev spaces for density. Anisotropic mesh adaptation is explored in \cite{jensen2018topology} to improve the description of solid domains in topology optimization of flow problems. Additionally, Wiker et al. \cite{wiker2007topology} explored the use of viscosity as a dependent parameter, providing examples of channels in a tree-shaped structure for pure Darcy problems and mixed Stokes–Darcy flow. The field of fluid flow TO has also been extended to three-phase interpolation models, considering fluid permeability through porous media and impenetrable inner walls using the solid isotropic material with penalization (SIMP) interpolation functions \cite{shen2018topology}. A Matlab implementation is presented in \cite{pereira2016fluid}, demonstrating stable low-order discretization of Stokes equations using polygonal finite elements. Parallel computations have been employed for large-scale  2D and 3D Stokes flow problems \cite{aage2008topology}. In \cite{liu2022marker}, a marker-and-cell method is introduced for large-scale optimization on GPU, utilizing a geometric multigrid preconditioner. Lastly, a phase field approach is introduced in \cite{garcke2015phase} for shape and topology optimization in Stokes flow, providing a well-posed problem in a diffuse interface setting. 

\subsection{Multiscale Topology Optimization}
\label{sec: multi_TO}
The methods discussed above result in a single-scale design where all design features are of the same length-scale. As discussed earlier, for problems with additional constraints, one must resort to  multi-scale TO (MTO), where one designs optimal microstructures in each finite-element cell while simultaneously solving the global flow problem \cite{wu2019topology}. Several MTO techniques have been proposed for structural and thermal problems. In \cite{allaire1997shape, allaire1993optimal}, the authors introduced techniques aimed at finding designs with optimal constitutive properties under structural and thermal loads. However, these theoretically optimal designs often lead to extremely length scales, posing manufacturing challenges \cite{groen2018homogenization, wu2021topology}. Additionally, their applicability to fluid problems remains unexplored. A more common approach is to compute optimal microstructures in each cell \cite{wu2019topology, coelho2008hierarchical, xia2014concurrent}. While offering a broad design freedom and applicability to various physical phenomena, including fluid flow \cite{wu2019topology}, these methods tend to be computationally expensive \cite{wu2021topology} since one must carry out homogenization of the evolving microstructures during each step of the optimization process \cite{zhou2008design}. 

\subsection{Variations of MTO}
\label{sec: data_TO}

To tackle the computational challenges of classic MTO, other techniques have been proposed. For example, graded-MTO (GMTO) \cite{nguyen2021multiscale, zhao2022stress, zheng2021data, wang2021data, wang2022data, watts2019simple, white2019multiscale,wang2017multiscale} employs graded variations of pre-selected microstructures. This allows for pre-computation of microstructural properties through offline homogenization before optimization \cite{chandrasekhar2023graded, li2019design}.  
One major limitation of GMTO is that the microstructural shape must be pre-selected prior to optimization. Thus,  new microstructure shapes cannot be discovered during optimization. Moreover, these pre-selected shapes are typically graded using a single parameter, which further restricts the variety of microstructures that are generated. To address the aforementioned challenge, a microstructure blending-based multiscale approach has been proposed in \cite{chan2022remixing}, which can generate new classes of microstructures. However, the approach requires supplementary parameters beyond the conventional shape parameters. Moreover, the blending process requires additional steps to impose bounds on the blending operation, to prevent any distorted or invalid shapes in the resulting microstructures.  In \cite{padhy2023fluto}, a set of microstructures were pre-selected, and their size/orientation were optimized. While this  slightly increased the design space, it is still limiting and leads to undesirable mixing of microstructures within each cell.

\subsection{Contributions}
\label{sec:intro_contribution}

In \cref{sec:method}, we propose an alternate and efficient MTO method that uses VAE in combination with super-shapes to compute fluid designs with low dissipation, and desired contact area. In \cref{sec:expts}, we demonstrate, using numerical experiments, that this significantly increase in design space. Conclusions and future work are discussed in \cref{sec:conclusion}.

\section{Proposed Method}
\label{sec:method}
\subsection{Assumptions and Strategy}
\label{sec:Fluid_eqs}
Consider a design domain $\Omega^0$ with prescribed flow boundary conditions as illustrated earlier in \cref{fig:Fluid_MTO}(a). The objective is to compute a multiscale design that minimizes the dissipated power subject to a total contact-area (i.e., perimeter in 2D) constraint. We will assume that it is a low-Reynolds flow, and the  fluid is  incompressible, i.e., the fluid is governed by Stokes equation:
\begin{subequations}
\begin{align}
    -2\nabla.[\mu\bm{\epsilon(u)}]+\bm{C_{eff}^{-1}}.\bm{u}+\nabla p= 0 \text{ in  $\Omega^0$} \\
    \nabla.\bm{u}=0  \text{ in  $\Omega^0$}\\
    \bm{u}= \bm{\mathcal{G}} \text{ over }  \partial\Omega^0
    \end{align}
    \label{eq:fluidGovnEq}
\end{subequations}
where $\bm{u}$ and p are the velocity vector and pressure of the fluid,  $\bm{\epsilon(u)} = (\bm{\nabla} \bm{u} + \bm{\nabla^T}\bm{u})/2$ represents the rate-of-strain tensor, $\bm{C_{eff}^{-1}}$ denotes the inverse of effective permeability tensor \textcolor{blue}{\cite{andreasen2011multiscale}} which penalizes the fluid flow in the design domain (for more details see \cref{sec:off_hom}). The  viscosity $\mu$ and mass density are assumed to be unity; $\bm{\mathcal{G}}$ is velocity field imposed on $\partial\Omega^0$.

The overall strategy is to discretize the domain into finite element cells, and dynamically create optimal microstructures in each cell from a single family of parameterized super-shapes, discussed next.

\subsection{Super-Shapes}
 Super-shapes, also known as Gielis curves, were introduced by Gielis in 2003 \cite{gielis2003generic} as an extension of super-quadrics. Unlike super-quadrics that utilize only two parameters, super-shapes incorporate six parameters namely, size: $a$ and $b$, order of rotational symmetry: $m$, and curvature: $n_{1}$, $n_{2}$, and $n_{3}$, where all numbers are assumed to be positive real. Using these six parameters, the super-shape boundary is defined  by the set of points:
 
\begin{equation}
    (x,y) = \big(r(\alpha)\cos(\alpha), r(\alpha)\sin(\alpha)\big)\; ; \; 0 \leq \alpha \leq 2\pi
    \label{eq:supershape_parameteric_form}
\end{equation}

where,
\begin{equation}
r(\alpha) = \left[ \left| \frac{1}{a} \cos \left( \frac{m \alpha}{4} \right) \right|^{n_{2}} + \left| \frac{1}{b} \sin \left( \frac{m \alpha}{4} \right) \right|^{n_{3}} \right]^{-\frac{1}{n_{1}}}
\label{eq:supershape_radius}
\end{equation}

\begin{figure}[H]
	\begin{center}
	\includegraphics[scale=0.3,trim={0 00 200 0} ]{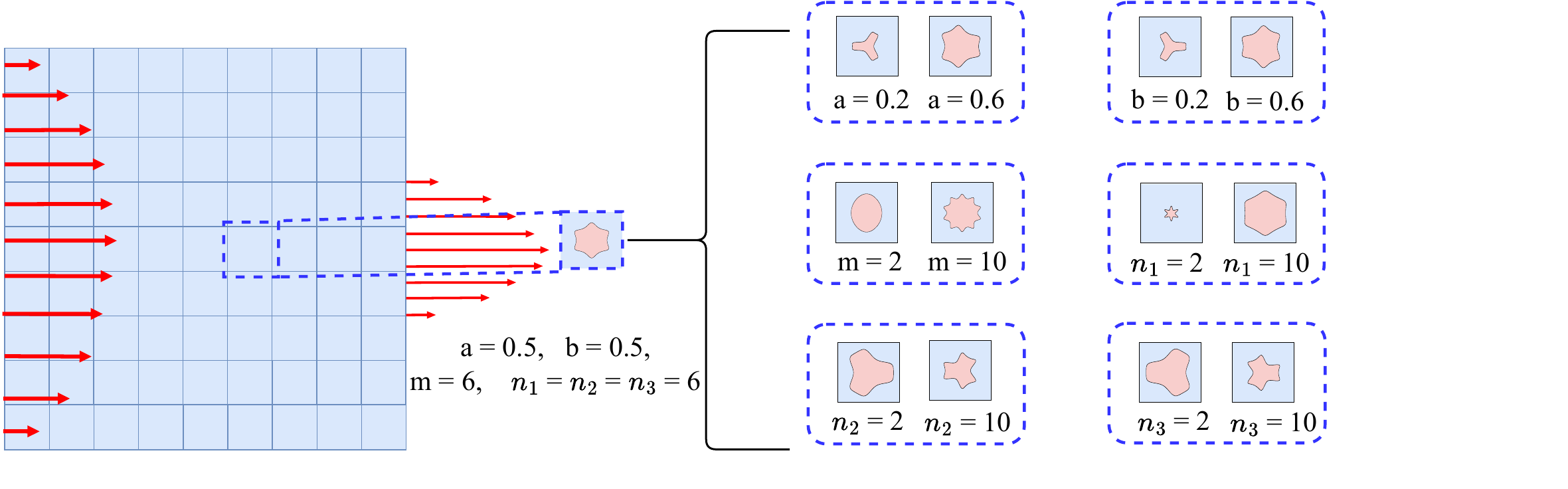}	\caption{A variety of microstructure generated using supershape parameters. } 
	\label{fig:problem_formulation}
	\end{center}
 \end{figure}

By varying these six parameters, a variety of shapes can be obtained as illustrated in  \cref {fig:problem_formulation}. 
 
\begin{figure}[H]
	\begin{center}
	\includegraphics[scale=0.45,trim={0 0 0 0} ]{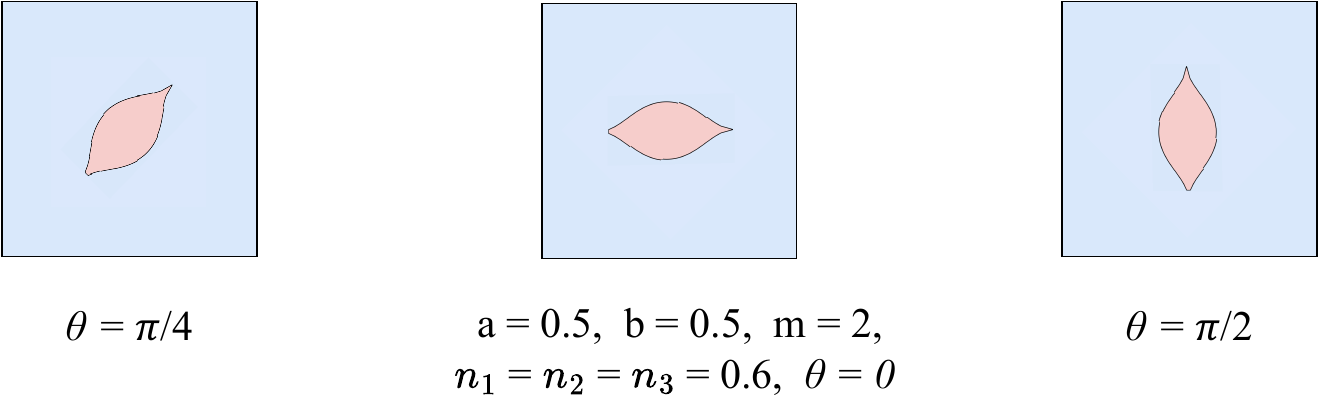}	
    \caption{A fish-shaped microstructure oriented at various angles.} 
	\label{fig:fish_body}
	\end{center}
 \end{figure}

In addition to varying these parameters, we allow the shapes be oriented with respect to the x-axis, using a orientation parameter $\theta$; see \cref {fig:fish_body}. 

The proposed strategy is to find optimal super-shape parameters, and orientation, within each cell that minimize the overall dissipated power, subject to a contact-area. A naive approach would entail computing the homogenized constitutive tensors of evolving super-shapes in each cell, during each step of the optimization process. This is once again computationally intractable. Instead,  we propose an off-line strategy where a finite set of super-shapes are analyzed, and their characteristics are captured using a  variational auto-encoder (VAE). The resulting decoder (and latent space) is then used for efficient multi-scale optimization. In the remaining sections, the proposed strategy is discussed in detail.

\subsection{Offline Computation}
\label{sec:off_hom}

\begin{figure}[H]
 	\begin{center}
	\includegraphics[scale=0.35,trim={0 0 0 0} ]{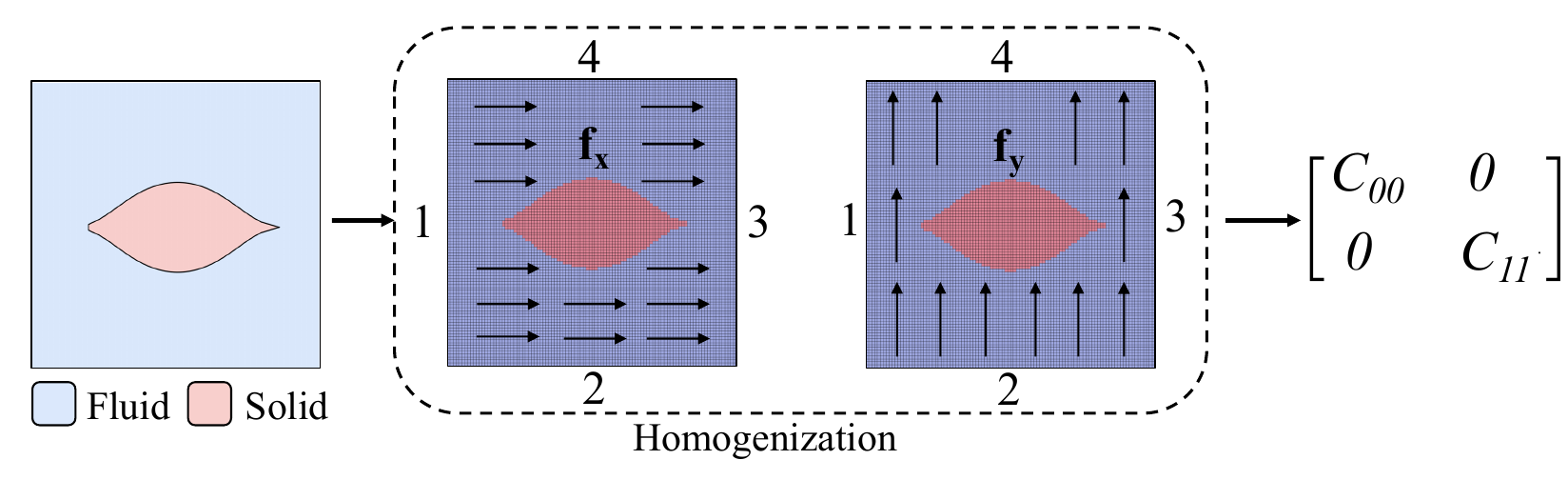}	\caption{Offline homogenization.}
	\label{fig:homogenization}
	\end{center}
 \end{figure}

We now describe the process for computing the permeability of any super-shape microstructure. Consider a generic super-shape with a given set of parameters $\bm{M} = \{a,b,m,n_1,n_2,n_3\}$ within a unit cell, as in  \cref{fig:homogenization}. The contact-area (i.e., perimeter) and volume fraction are first computed by discretizing the boundary and then computing these geometric quantities using the $\textit{shapely}$ library \cite{shapely_software}.  To compute the $2 \times 2 $ permeability tensor $\bm{C}$, the domain is discretized into a mesh of 150 × 150 elements.  Then, two Stokes flow problems are solved, subject to unit body forces $f_x = 1$ and $f_y = 1$, as illustrated in \cref{fig:homogenization}. The boundary conditions involve coupling boundaries 1 and 3 through periodic conditions for velocity and pressure, as well as coupling boundaries 2 and 4 in a similar manner; see \cite{andreassen2014determine}. The velocities obtained from solving the problem with $f_x = 1$ are denoted as $u_0(x,y)$ and $v_0(x,y)$, while those obtained from solving the problem with $f_y = 1$ are denoted as $u_1(x,y)$ and $v_1(x,y)$. Since $u_1(x,y)$ and $v_0(x,y)$ are nearly orthogonal to the bulk flow directions,   the off-diagonal terms of the permeability tensor are three orders of magnitude smaller than the diagonal terms, and can be neglected \cite{wang2013full}. Thus, the permeability tensor $\bm{C}$ is computed as follows \cite{andreassen2014determine, lang2014permeability, vianna2020computing}: 

\begin{equation}
    \bm{C} =  
    \begin{bmatrix} C_{00} & 0 \\
    0  & C_{11}
    \end{bmatrix} = 
    \frac{1}{|V|} \begin{bmatrix} \int\limits_{V}u_{0}dV &   0 \\
     0 &  \int\limits_{V}v_{1}dV \end{bmatrix}
    \label{eq:CMatrix}
    \end{equation}

where the volume $V$ of the unit cell is unity. Note that the the orientation is taken into account through the following tensor operation to determine the effective permeability tensor \cite{lang2014permeability}:
\begin{equation}
    \bm{C}_{eff} =  \begin{bmatrix} \cos(\theta) & -\sin(\theta) \\ \sin(\theta) & \cos(\theta) \end{bmatrix}
    \begin{bmatrix} C_{00} & 0 \\
    0 & C_{11} \end{bmatrix}
    \begin{bmatrix} \cos(\theta) & -\sin(\theta) \\ \sin(\theta) & \cos(\theta) \end{bmatrix}^T
    \label{eq:effectiveCMatrix}
\end{equation}

For numerical homogenization, a Brinkman penalization of zero is applied for the fluid phase and $10^6$ for the solid phase. For further details on the numerical homogenization methodology employed in this study, please see \cite{andreassen2014determine}.  A random set of 7000 samples of super-shapes are analyzed using the above process, where parameter instances are generated using a uniform random distribution  \cite{oliphant2006guide} as follows:  $0.05 \le a, b \le 0.75 $, $1 \le m \le 22$ and $0.5 \le n_{1}, n_{2}, n_{3}  \le 10$. The results from the offline computation are then analyzed using variational autoencoders, discussed next.
 
\subsection{Variational Auto-Encoders}
\label{sec:method_vae}

Variational auto-encoders (VAEs) are a type of generative model that leverages probabilistic encoding and decoding techniques to compress input data into a lower-dimensional latent space \cite{kingma2019introduction, kingma2013auto, doersch2016tutorial}. One of the key advantages of VAEs, as opposed to other encoding methods, is their ability to generate new samples that resemble the original input data \cite{doersch2016tutorial}. For instance, VAEs have been successfully employed to generate novel microstructures  from image databases \cite{wang2020deep}. Another important feature of VAEs is the creation of continuous and differentiable latent space. This allows for gradient-based optimization, enabling efficient exploration of the latent space. This is particularly valuable in applications such as reliability based TO \cite{gladstone2021robust}. Finally, unlike linear dimensionality reduction techniques such as principal component analysis (PCA), VAEs can learn complex non-linear relationships between the input and the reduced dimensionsal space \cite{heaton2018ian}. 

\begin{figure}[H]
 	\begin{center}
	\includegraphics[ scale= 0.9, trim={0 0 0 0} ]{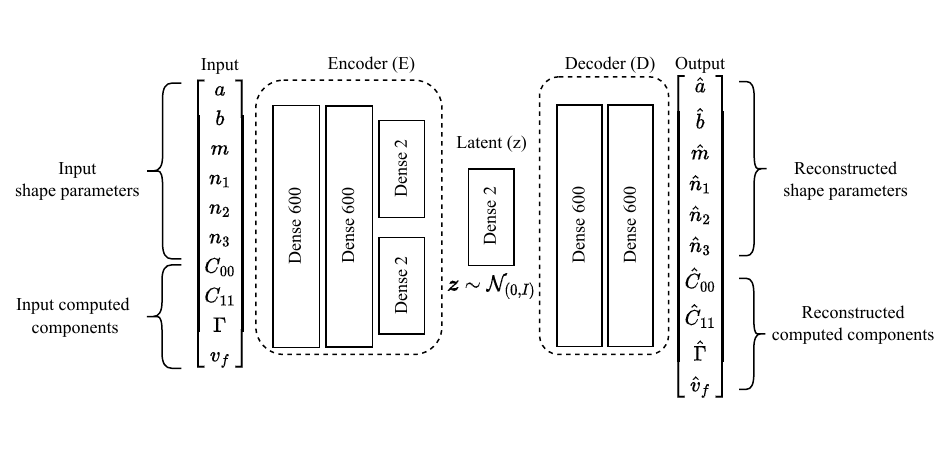}
 \caption{Proposed VAE network.}
	\label{fig:vae_net}
	\end{center}
 \end{figure}

The proposed VAE architecture, depicted in \cref{fig:vae_net}, consists of several essential components:
\begin{enumerate}
    \item Firstly, the input $\bm{\Psi}$ is ten-dimensional representing six shape parameters $\bm{M}$, two permeability components ($C_{00}$ and $C_{11}$), contact area $\Gamma$ and volume fraction $v_f$.
    \item  The encoder $E$, following along \cite{higgins2016betaVAE}, is a fully-connected network consisting of two hidden layers, each containing 600 neurons with a LeakyReLU activation functions \cite{schmidhuber2015deep}. 
    \item The proposed VAE uses a two-dimensional latent space ($z_{1}$ and $z_{2}$).
    \item Additionally, a decoder  $D$ is constructed with two hidden layers, each containing 600 neurons. 
    \item The output consists of the same ten properties: shape parameters $\hat{\bm{M}}$, permeability components $\hat{\bm{C}}$ $\equiv$ ($\hat{C}_{00}$ and $\hat{C}_{11}$), contact area $\hat{{\Gamma}}$ and volume fraction $\hat{v}_f$ . They can be combined as $\bm{\hat{\Psi}}$ $\equiv$ ($\hat{a}$, $\hat{b}$, $\hat{m}$, $\hat{n}_{1}$, $\hat{n}_{2}$, $\hat{n}_{3}$, $\hat{C}_{00}$, $\hat{C}_{11}$, $\hat{\Gamma}$, $\hat{v}_f$).  
\end{enumerate}

Note that the reconstruction will not be exact. The VAE is trained to minimize the difference between the output and input \cite{rautela2022towards}. This involves optimizing the weights associated with the encoder and decoder. Additionally, to enforce the latent space to approximate a Gaussian distribution, a KL divergence loss term is introduced \cite{kingma2019introduction}. Thus, the overall VAE loss function can be formulated as follows:
\begin{equation}
    L_v = ||\bm{\Psi} - \bm{\hat{\Psi}} ||_2 + \beta\text{KL}(z || \mathcal{N})
    \label{eq:vae_training_loss}
\end{equation}

Here, $\beta$ is set to  $10^{-7}$ \cite{rautela2022towards}. To achieve a stable convergence, the geometric parameters, contact area and volume fraction are normalized linearly between $0$ and $1$, while the permeability components are scaled logarithmically due to  significant variation in magnitude. 



\subsection{Latent Space}
\label{sec:method_vae_latentSpace}

\begin{figure}[H]
 	\begin{center}
	\includegraphics[scale= 0.35, trim={0 0 0 0} ]{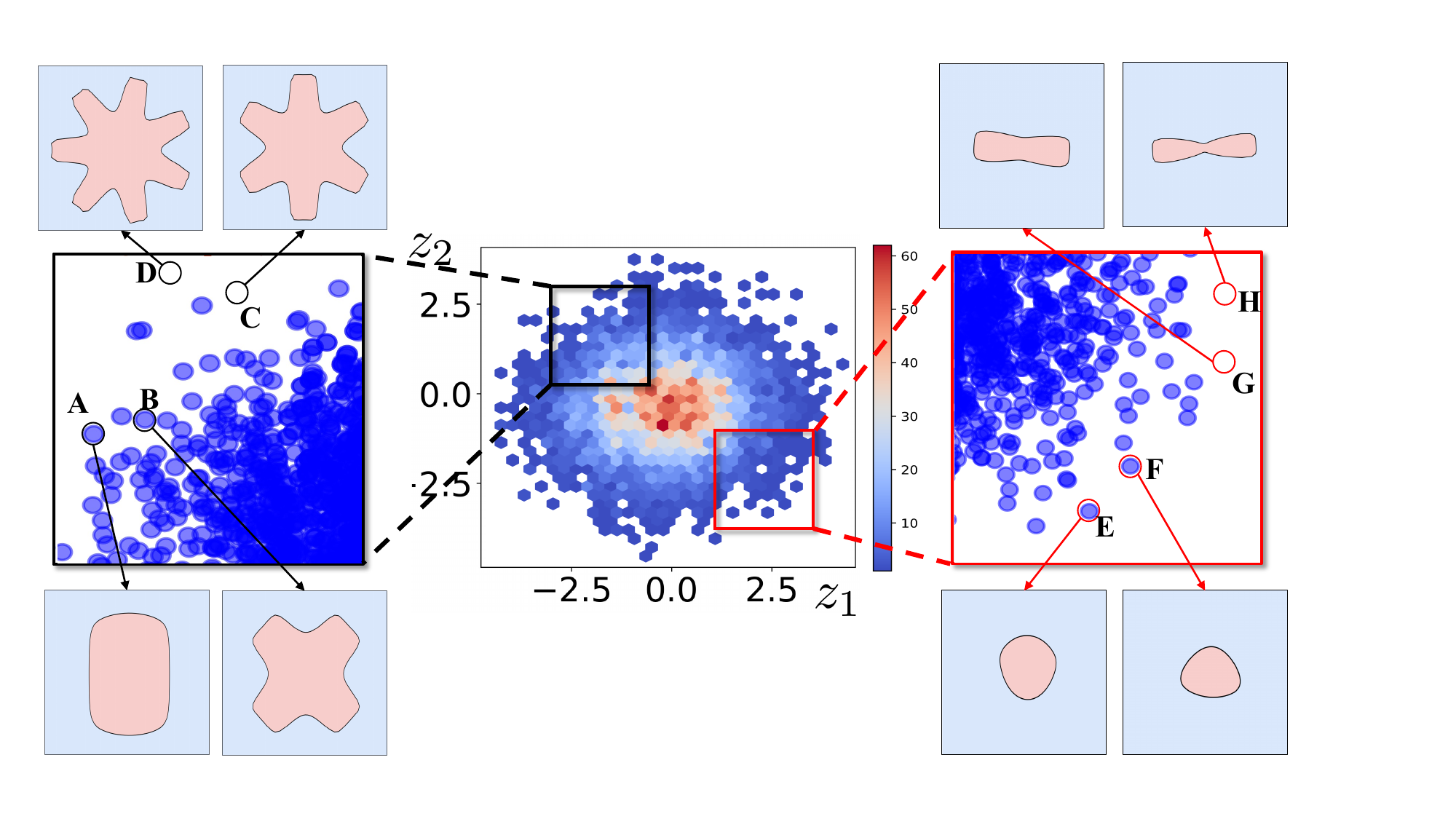}	\caption{The latent space density distribution and scattered plots in the insets reveal both microstructures existing in the dataset and new microstructures generated by the VAE that are not originally present in the dataset.}
	\label{fig:latent_space_with_shapes}
	\end{center}
 \end{figure}

Once the latent space has been constructed, the trained decoder $D^*$ can be used to generate super-shape parameters and properties via $\hat{\Psi}$ = $D^*(z_1, z_2)$ for all points within the latent space. 
The generated latent space has the following features:

\begin{enumerate}
    \item \textbf{Generation of new microstructures}: Although the data-set used to train the decoder is discrete, the resulting latent space is continuous. This continuous representation facilitates a meaningful exploration of microstructure configurations throughout the latent space. For example in \cref{fig:latent_space_with_shapes}, while A, B, E, and F depict points present in the data-set, points C, D, G, and H are generated by the trained decoder, with corresponding microstructures. 
    
    \item \textbf{Differentiable Latent Space}: The latent space is differentiable in that derivatives such as $\frac{\partial {\hat{\Psi}}}{\partial \bm{z_1}}$, can be computed analytically using back-propagation. This enables gradient-based optimization. 
    
    \item \textbf{Compact representation and computational efficiency}: One of the main advantages of using VAE for latent space generation is the reduction in dimensionality it offers, compacting the original data ($\bm{\Psi}$) from 10 dimensions to a 2-dimensional latent space $(z_1, z_2)$. This reduces the number of design variables and thereby reducing the computational burden. 
\end{enumerate}

\subsection{Global Fluid Flow Analysis}
\label{sec:Fluid_FEA}

We are now ready to address global fluid flow analysis. Here,  a quadrilateral Q2-Q1 (quadratic velocity/linear pressure) element belonging to the class of the Taylor-Hood elements is used. The elemental stiffness matrix $\bm{K_e}$ and degrees of freedom vector $\bm{S_e}$ for the governing equation (see \cref{sec:Fluid_eqs}) are given by (see \cite{pereira2016fluid} for details): 
\begin{equation}
\bm{K_e} = 
\begin{bmatrix}
    \bm{A_e} && \bm{B_e} && \bm{0} \\
    \bm{B_e^T}  && \bm{0} && \bm{h_e}\\
    \bm{0}  && \bm{h_e^T} && \bm{0}
    \end{bmatrix} \;, \; \bm{S_e}=
\begin{bmatrix}
    \bm{U_e}  \\
    \bm{P_e}  \\
    \bm{\lambda}
    \end{bmatrix} 
    \label{eq:StiffnessMatrix}
\end{equation}
where
\begin{subequations}
	\label{eq:stiffness_term_defn}
	   \begin{align}
	& \bm{A_e} = \bm{A_e^\mu}+\bm{C_{eff, e}}^{-1}\bm{A_e^\alpha}\\
	&\bm{[A_e^\mu]}_{ij} = \int_{\Omega_e}2\mu\bm{\epsilon(N_i)}:\bm{\epsilon(N_j)}d\Omega\\
     &\bm{[A_e^\alpha]}_{ij} = \int_{\Omega_e}\bm{N_i}\bm{N_j}d\Omega\\
     &\bm{[B_e]}_{ij} = \int_{\Omega_e}\bm{L_j}\nabla.\bm{ N_i}d\Omega\\
     &\bm{[h_e]}_{i} = \int_{\Omega_e}\bm{L_i}d\Omega
	\end{align}
\end{subequations}
Here, $\bm{N_i}$ and $\bm{L_i}$ are the velocity and pressure basis functions, \bm{$U_e$} and \bm{$P_e$} represent elemental velocity and pressure degrees of freedom respectively and $\bm{C_{eff,e}}  $ is the design dependent effective element permeability matrix (\cref{sec:off_hom}).  In order to uniquely define the pressure field, a zero mean condition is enforced by $\bm{h_e^T}P = 0$ (for details see \cite{pereira2016fluid}). The individual elemental $\bm{K_e}$ and $\bm{S_e}$ matrices are assembled to construct the global stiffness matrix $\bm{K}$ and degrees of vector $\bm{S}$ respectively. We then solve the equation $\bm{K}$$\bm{S}$ = $\bm{f}$, wherein the vector $ \bm{f}$ represents the boundary conditions applied. This solution determines the unknown degrees of freedom in $\bm{S}$.

\subsection{Design Variables, Objective and Constraints}
Finally, the optimization framework comprises of the following:
\label{sec:Opt_Cons}

\textbf{Design Variables:}
The  design variables associated with each element are denoted by $\bm{\zeta_{e}} = \{{z_{1, e}, z_{2, e}, \theta_{e}}\}$, where $z_{1, e}$ and $z_{2, e}$ are the two latent space variables, and $\theta_{e}$  is the orientation of the super-shape. The values $z_{1}$ and $z_{2}$ are constrained to lie within $[-3,3]$, and the orientation parameter is constrained as $0 \leq \theta \leq 2\pi $. The entire set of design variables is denoted by $\bm{\overline{\zeta}} = \{\bm{\zeta_{1}}, \bm{\zeta_{2}}, ...., \bm{\zeta_{N_{e}}}\}$.

\textbf{Objective:} 
The objective is to minimize the dissipated power given by \cite{borrvall2003topology, pereira2016fluid}:
\begin{equation}
    \label{eq:min_obj}
 J(\overline {\bm{\zeta}}) = \sum\limits_{e=1}^{N_e}\frac{1}{2}\bm{U_e}^T\bm{[A_e^\mu+A_e^\alpha C_{eff,e}^{-1}]U_e}
\end{equation}

\textbf{Contact-Area Constraint:} The contact area $\hat{\Gamma}_{e}$ of each microstructure is reconstructed using the decoder, and the following global constraint is imposed:
\begin{equation}
    \label{eq:contact_cons}
    g_{\Gamma} (\overline {\bm{\zeta}}) \equiv 1 - \frac{\sum\limits_{e=1}^{N_e} \hat{\Gamma}_{e} }{\Gamma _{min}} \leq 0
\end{equation}

where ${\Gamma _{min}}$ is the lower bound on the contact area and $N_e$ represents the number of elements in the design domain.

\textbf{Volume Constraint:} Instead of imposing a contact area constraint, one can impose a volume constraint: 
\begin{equation}
    \label{eq:vol_cons}
    g_{V} (\overline {\bm{\zeta}}) \equiv \frac{\sum\limits_{e=1}^{N_e}\hat{v}_{f, e}} { N_e v_{max}} - 1 \leq 0
\end{equation}
where, $\hat{v}_{f, e}$ is the fluid volume fraction, and $v_{max}$ is the upper bound on the volume fraction.

\subsection{Multiscale Optimization Problem}
\label{sec:method_TO}

Consequently, one can pose the multiscale problem in a finite-element setting as:
\begin{subequations}
	\label{eq:optimization_base_Eqn}
	\begin{align}
		& \underset{\overline {\bm{\zeta}} = \{ \bm{\zeta}_1, \bm{\zeta}_2, \ldots \bm{\zeta}_{N_e} \}} {\text{minimize}}
		& J(\overline {\bm{\zeta}}) &= \sum\limits_{e=1}^{N_e}\frac{1}{2}\bm{U_e}^T\bm{[A_e^\mu+A_e^\alpha C_{eff, e}^{-1}]U_e} \label{eq:optimization_base_objective}\\
		& \text{subject to}
		&  \bm{K}(\overline {\bm{\zeta}})\bm{S} & = \bm{f}\label{eq:optimization_base_govnEq}\\
		& &  g_{\Gamma} (\overline {\bm{\zeta}}) & \equiv 1 - \frac{\sum\limits_{e=1}^{N_e} \hat{\Gamma}_{e} }{\Gamma _{min}}  \leq 0  \label{eq:optimization_base_perimCons} \\
      & & \text{(or)} \quad  g_{V} (\overline {\bm{\zeta}}) & \equiv \frac{\sum\limits_{e=1}^{N_e}\hat{v}_{f, e}} { N_e v_{max}} - 1 \leq 0  \label{eq:optimization_base_volCons} \\
		& &  -3 \leq z_{e,0}, z_{e,1} &\leq 3 \; , \; \forall e \; \label{optimization_base_boundConsRho} \\
		& & 0 \leq  \theta_e & \leq 2 \pi \; , \; \forall e \label{optimization_base_boundConsTheta}
	\end{align}
\end{subequations}
To solve the above optimization problem, optimization techniques such as method of moving asymptotes \cite{svanberg1987method} or optimality criteria \cite{rozvany2012structural} can be used. However, we use neural networks for optimization \cite{chandrasekhar2021tounn} due to the advantages discussed in the subsequent section.

\subsection{Optimization using a Neural Network}
\label{sec:method_TO_NN}
\begin{figure}
 	\begin{center}
	\includegraphics[scale=0.8,trim={0 0 0 0} ]{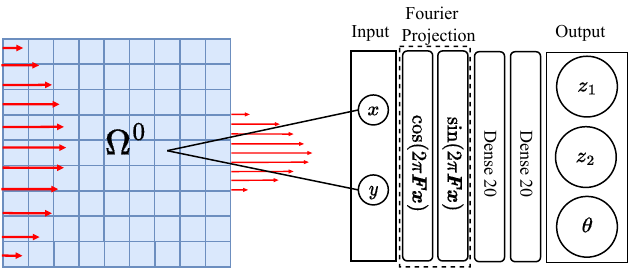}	
        \caption{Topology optimization network.}
	\label{fig:top_net}
	\end{center}
 \end{figure}

Neural networks (NN) are employed for global optimization as they inherently support automatic differentiation \cite{pyTorch}, which enables seamless gradient calculations. The proposed neural-network (NN) architecture for global optimization is illustrated in \cref{fig:top_net}, and it consists of the following entities:

\begin{enumerate}
\item \textbf{Input Layer}: The input to the NN are points $\bm{x} \in \mathbf{R}^2$ within the domain $\Omega^0$. Although these points can be arbitrary,  they correspond here to the center of the elements.

    \item \textbf{Fourier Projection}: The sampled points from the Euclidean domain are directed through a frequency space, associated with a frequency range $\bm{F}$. Prior research \cite{rahaman2019spectral, tancik2020fourier} indicate that implicit coordinate-based neural networks are biased to lower frequency components of the target signal. To address this issue and speed up convergence, a Fourier projection layer is integrated before the standard activation layers \cite{chandrasekhar2022approximate}.
        
    \item \textbf{Hidden Layers}: The hidden layers consist of a series of fully connected LeakyReLU activated neurons, LeakyReLU is a differentiable function, as opposed to ReLU, and is therefore preferred in this work \cite{maas2013rectifier}. 
    In particular, the neural network used here consists of two hidden layers, each activated with the LeakyReLU
    function, and each layer has 20 neurons.
	
	\item \textbf{Output Layer}: The output layer consists of $3$ neurons corresponding the design variables for each element $\bm{\zeta_e} = \{z_{1}(\bm{x}), z_{2}(\bm{x}), \theta(\bm{x})\}$. The output neurons are activated by a Sigmoid function  $\sigma(\cdot)$. The neurons associated with the latent space variables are scaled as $z_{i} \leftarrow -3 + 6 \sigma(z_{i}) $ to retrieve values in the range of a standard Gaussian Normal distribution. Further, the output neuron associated with the orientation is scaled as $\theta \leftarrow 2\pi \sigma (\theta)$. Thus, the box constraints in \cref{optimization_base_boundConsRho} and \cref{optimization_base_boundConsTheta} are not needed.
	
	\item \textbf{NN Design Variables}: The weights and bias associated with the NN, denoted by the $\bm{w}$, now become the primary design variables, i.e., we have  $z_{1}(\bm{x}; \bm{w}),$ $z_{2}(\bm{x}; \bm{w})$ and $\theta(\bm{x}; \bm{w})$.
    The weights in the network are initialized using Xavier weight initialization \cite{glorot2010understanding} with a seed value of 77.

    \item \textbf{Optimizer:} The Adam optimizer is used with a learning rate of $4\cdot10^{-3}$. The optimization process is set to run for a maximum of 300 iterations (epochs). To ensure convergence, the optimization monitors the change in loss  ($\Delta L_c^{*}$) (see \cref{eq:lossFunction}) with a threshold of $10^{-5}$.
 
\end{enumerate}

Thus,  \cref{eq:optimization_base_Eqn} reduces to:
\begin{subequations}
	\label{eq:optimization_nn_Eqn}
	\begin{align}
		& \underset{\bm{w}}{\text{minimize}}
		& &J(\bm{w}) \label{eq:optimization_nn_objective}\\
		& \text{subject to}
		& & \bm{K}(\bm{w})\bm{S} = \bm{f}\label{eq:optimization_nn_govnEq}\\
		& & & g_{\Gamma} (\bm{w}) \equiv  1 - \frac{\sum\limits_{e=1}^{N_e} \hat{\Gamma}_{e} (\bm{w})}{\Gamma_{min}}  \leq 0  \label{eq:optimization_nn_perimCons}\\
            & &  \text{(or)} \quad  g_{V} (\bm{w}) & \equiv   \frac{\sum\limits_{e=1}^{N_e} (\hat{v}_{f, e} (\bm{w}))}{ N_e  v_{max}}-1  \leq 0 \label{eq:optimization_nn_volCons}
	\end{align}
\end{subequations}

Since neural networks are designed to minimize an unconstrained loss function, we convert the constrained minimization problem into a loss function minimization by employing the penalty scheme  \cite{wright2006numerical}. Specifically, the loss function is defined as:
\begin{equation}
    L_T(\bm{w}) = \frac{J(\bm{w})}{J^0} + \gamma g(\bm{w})^2
    \label{eq:lossFunction}
\end{equation}
where the parameter $\gamma$ is updated during each iteration, making the enforcement of the constraint stricter as the optimization progresses. The constraint penalty in the current framework starts with an initial value of $\gamma$ = 1. and is incremented by $\Delta \gamma = 0.1$ after every epoch. 

\begin{figure}[H]
 	\begin{center}
	\includegraphics[width=100mm,scale=1,trim={0 0 0 0} ]{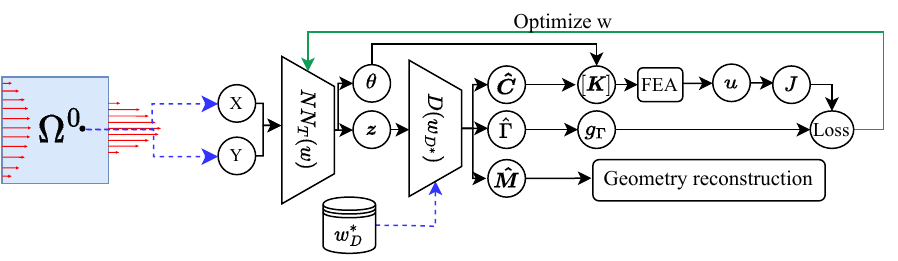}	
 \caption{Optimization flowchart.}
	\label{fig:opt_flow}
	\end{center}
 \end{figure}

Thus, the overall framework is illustrated in \cref{fig:opt_flow}.

\subsection{Sensitivity Analysis}
\label{sec:method_sensitivity}

A critical ingredient in gradient-based optimization is the sensitivity, i.e., derivative, of the objective and constraint(s) with respect to the optimization parameters.  Typically the sensitivity analysis is carried out manually. For example,  the derivatives of the objective function is typically expressed as follows where each term is computed manually:

\begin{equation}
    \frac{\partial L_T}{\partial \bm{w}} = \big[
    \frac{\partial L_T}{\partial J} 
    \frac{\partial J}{\partial \bm{u}} 
    \frac{\partial \bm{u}}{\partial \bm{K}} \big(
    \frac{\partial \bm{K}}{\partial \hat{\bm{C}}}  
    \frac{\partial \hat{\bm{C}}}{\partial \bm{w_D^*}} 
    \frac{\partial \bm{w_D^*}}{\partial \bm{z}}
    \frac{\partial\bm{z}}{\partial \bm{w}} + 
    \frac{\partial \bm{K}}{\partial \bm{\theta}}  
    \frac{\partial \bm{\theta}}{\partial \bm{w}}
    \big) + 
    \frac{\partial L_T}{\partial g_{\Gamma}}
    \frac{\partial g_{\Gamma}}{\partial \hat{\bm{\Gamma}}} 
    \frac{\partial \hat{\bm{\Gamma}}}{\partial \bm{w_D^*}} 
    \frac{\partial \bm{w_D^*}}{\partial \bm{z}}
    \frac{\partial\bm{z}}{\partial \bm{w}}
    \big] 
    \label{eq:sens_objective}
\end{equation}

This can be laborious and error-prone, especially for non-trivial objectives. Here, by expressing all our computations including computing the permeability tensors, stiffness matrix, FEA, objectives and constraints in PyTorch \cite{NEURIPS2019_9015}, we use the NN's automatic differentiation (AD) capabilities to completely automate this step  \cite{chandrasekhar2021auto}. In other words, only the forward expressions need to be defined, and all required derivatives are computed to machine precision by PyTorch computing library.

\subsection{Algorithms }
\label{sec:Algorithm}

The three algorithms used here are summarized in  \cref{alg:DG}, \cref{alg:ED} and \cref{alg:FTO}. In the first algorithm, the primary objective is to generate a set of microstructures and their properties  $\Psi$.

\begin{algorithm}[]
    \caption{Generate Microstructure properties data-set}
    \label{alg:DG}
    \begin{algorithmic}[1]
        \Procedure {DataGen}{}  
        \State $ \bm{M} \rightarrow \Gamma, v_f,  C_{00},  C_{11}$ \Comment{Offline Computation \cref{sec:off_hom}} 
        \State $  \bm{M}, \Gamma, v_f,  C_{00},  C_{11} \rightarrow \Psi$ \Comment{Data-set creation} 
    \EndProcedure    \Comment{Output: Microstructure data-set} 
    \end{algorithmic}
\end{algorithm}

In \cref{alg:ED}, using $\Psi$, the VAE is trained to produce a lower-dimensional latent space.  

\begin{algorithm}[]
    \caption{Encode Microstructure properties}
    \label{alg:ED}
    \begin{algorithmic}[1]
        \Procedure {MstrEncode}{$\Psi$, E, D}  \Comment{Input: Training data, encoder and decoder} 
        \State  epoch = 0 \Comment{iteration counter} \label{alg:epoch_count}
        \Repeat \Comment{VAE training}
        \State $ E(\Psi) \rightarrow \bm{z}$ \Comment{Forward prop.encoder} 
        \State $ D(\bm{z}) \rightarrow  \hat{\Psi}$ \Comment{Forward prop.decoder} 
        \State $ \{\Psi,\hat{\Psi}, \bm{z}\} \rightarrow  L_V$ \Comment{VAE loss \cref{eq:vae_training_loss}} 
        \State $\bm{w} , \nabla L_V \rightarrow \bm{w} $ \Comment{Adam optimizer step; update weights}\label{alg:AdamStep_encoder}
        \State $\text{epoch}++$
        \Until{ $|| \Delta L_V || < \Delta \hat{L}_V$ or epoch < max\_epoch} \Comment{check for convergence}
    \EndProcedure    \Comment{Output: Trained Decoder}
    \end{algorithmic}
\end{algorithm}

Once the training is complete, the encoder $E$ is discarded, and the decoder $D$ is retained. The  main optimization algorithm is summarized in \cref{alg:FTO}. First, the domain $\Omega^0$ is discretized for finite element analysis, and  the stiffness matrix components are computed (line 3). The mesh is sampled at the center of each element (line 4); these serve as inputs to the NN. The penalty parameter $\gamma$ and NN weights $\bm{w}$ are initialized \cite{glorot2010understanding} (line 5).

In the main iteration, the design variables $\overline{\bm{\zeta}}$ are computed using the NN (line 7). Then the latent space variable serve as input to the trained decoder $D^{*}$, followed by the computation of the microstructural geometric properties $\hat{\bm{M}}$, permeability components $\hat{\bm{C}}$ and contact area $\hat{{\Gamma}}$ (or, alternately, the volume fraction) for each element (line 8). The reconstructed permeability components from decoder along with the orientation from the NN is used to calculate the effective permeability tensor (line 9). The effective permeability is used to construct the stiffness matrix and to solve for the velocity and pressure  (lines 10-11). Then the objective and contact area (or volume) constraint are computed (lines 12 - 13), leading to the loss function (line 14). The  sensitivities are computed in an automated fashion (line 15). The weights $\bm{w}$ are then updated using Adam optimization scheme (line 16).  Finally the penalty parameters are updated (line 17). The process is repeated until termination, i.e., until the relative change in loss is below a certain threshold or the iterations exceed a maximum value.

\begin{algorithm}[]
	\caption{Fluid Topology Optimization}
	\label{alg:FTO}
	\begin{algorithmic}[1]
		\Procedure{TopOpt}{$\Omega^0$, BC,   $\Gamma_{min}$, $D^*$}  \Comment{Input: Design domain, boundary conditions, area constraint, and trained decoder} 
  
		\State $\Omega^0 \rightarrow \Omega^0_h$ \Comment{discretize domain for FE} \label{alg:domainDiscretize}\cref{sec:Fluid_FEA}
		
		\State $\Omega^0_h \rightarrow \bm{A^\mu},\bm{A^\alpha},\bm{B}, \bm{h}$ \Comment{ compute stiffness matrices} \label{alg:stiffnessTemplates}\cref{eq:stiffness_term_defn}
		
		\State $\bm{x} = \{x_e,y_e\}_{e \in \Omega^0_h} \quad \bm{x} \in \mathbb{R}^{n_e \times 2}$ \Comment{elem centers; NN input} \label{alg:elemCenterComp}

		\State  epoch = 0; $\gamma = \gamma_0$; $\bm{w} = \bm{w}_0$ \Comment{initialization} \label{alg:initalizationParams}
		
		\Repeat \Comment{optimization (Training)}
		
		\State $NN(\bm{x} ; \bm{w}) \rightarrow \overline{\bm{z}}(\bm{x}), \overline{\bm{\theta}}(\bm{x})$ \Comment{fwd prop through NN \cref{sec:method_TO_NN}} \label{alg:fwdPropNN}
  
		\State $D^{*}(\overline{\bm{z}}(\bm{x} ; \bm{w})) \rightarrow \hat{\bm{M}}({\bm{x}}),\hat{\bm{C}} \equiv (\hat{{C}}_{00}({\bm{x}}), \hat{{C}}_{11}({\bm{x}})), \hat{{\Gamma}}(\bm{x})$ \Comment{fwd prop through $D^*$} \label{alg:fwdPropD}
  
		\State $\hat{{C}}_{00}({\bm{x}}), \hat{{C}}_{11}({\bm{x}}), \theta({\bm{x}}) \rightarrow \bm{C}_{eff}(\bm{x}) $ \Comment{Effective permeability tensor \cref{eq:effectiveCMatrix} } \label{alg:PermeabilityTensorCompute}

		\State$ \bm{C}_{eff}(\bm{x}) \rightarrow \bm{K}, \bm{f} $ \Comment{ Stiffness matrix } \label{alg:stiffnessMatrix}
  \cref{eq:StiffnessMatrix}
				
		\State $\bm{K}, \bm{f}  \rightarrow \bm{S}$ \Comment{ solve \cref{eq:optimization_nn_govnEq}} \label{alg:feSolve}
		
		\State$\bm{K}, \bm{S} \rightarrow J$ \Comment{Objective, \cref{eq:optimization_nn_objective}} \label{alg:objectiveComp}

		\State $ \hat{{\Gamma}}, \Gamma_{min}  \rightarrow g_{\Gamma}$ \Comment{Contact area constraint }
  \cref{eq:contact_cons}\label{alg:perimCons}
		
		\State $J, g_{\Gamma} \rightarrow L$ \Comment{loss from \cref{eq:lossFunction}} \label{alg:lossCompute}
		
		\State $AD(L, \bm{w}) \rightarrow \nabla L $ \Comment{sensitivity analysis via Auto. Diff} \label{alg:autoDiff}
			 
		\State $\bm{w} , \nabla L \rightarrow \bm{w} $ \Comment{Adam optimizer step}\label{alg:AdamStep_topopt}
		
		\State $  \gamma + \Delta \gamma \rightarrow \gamma$ \Comment {increment penalty} \label{alg:OptPenaltyUpdate}
		
		\State $\text{epoch}++$
		
		\Until{ $|| \Delta L || < \Delta L_c^*$ or epoch < max\_epoch} \Comment{check for convergence}
		
		\EndProcedure
	\end{algorithmic}
\end{algorithm}

\section{Numerical Experiments}
\label{sec:expts}

In this section, we conduct several experiments to demonstrate the proposed framework. All experiments were conducted on a MacBook M2 Air, using the PyTorch library \cite{pyTorch} in Python.

\subsection{Ideal Microstructure Selection}
\label{sec:single_mstr}

\begin{figure}[H]
 	\begin{center}
	\includegraphics[width=60mm,scale=1.,trim={0 0 0 0} ]{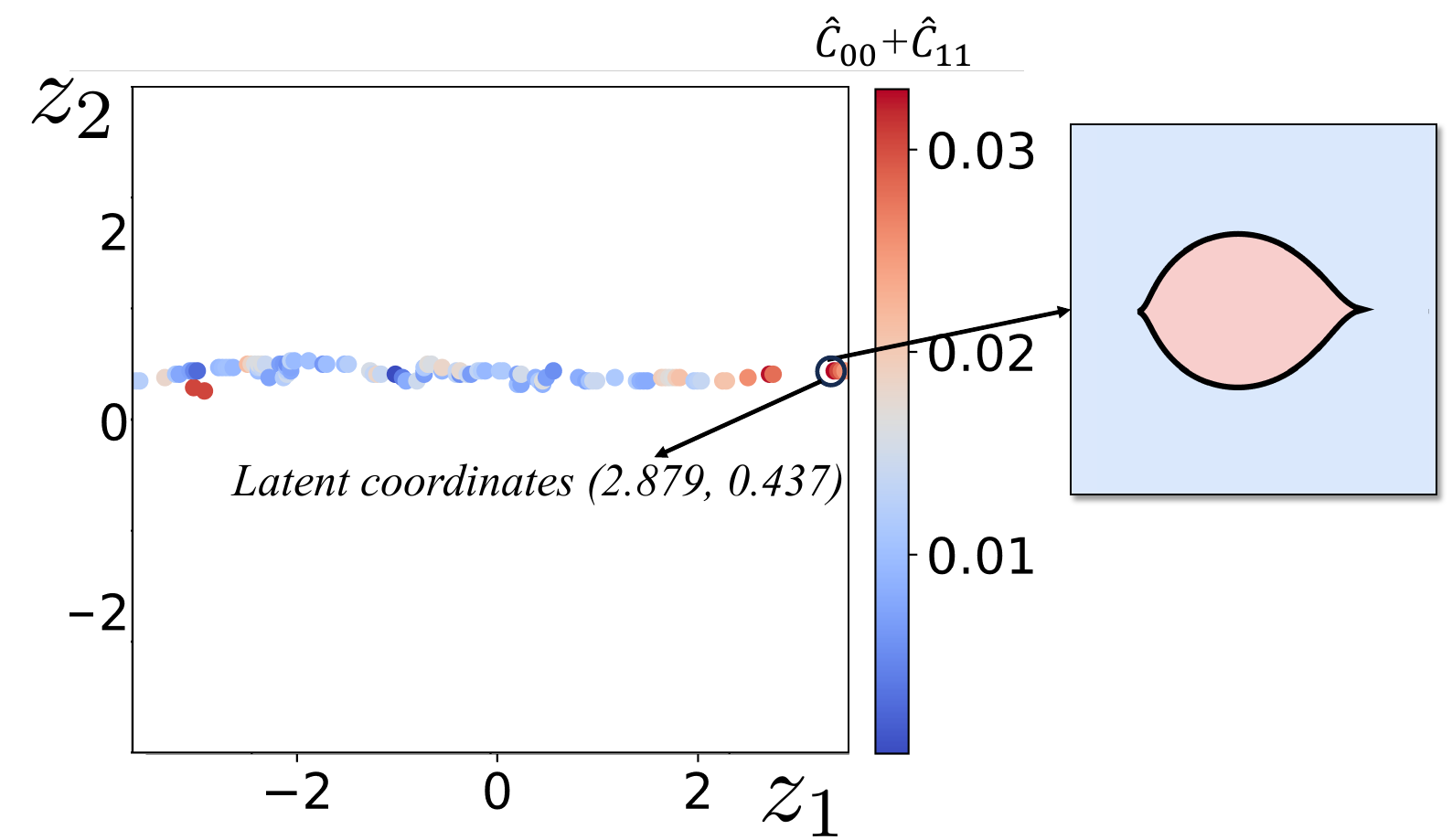}	\caption{Latent space coordinates of microstructures with solid volume fraction of approximately 0.25. }
	\label{fig:latent_fish}
	\end{center}
 \end{figure}

In this experiment, our objective is to identify a microstructure with a solid volume fraction of approximately $0.25$, with highest permeability. Towards this end, the latent space is uniformly sampled at $200 \times 200$ points using the decoder. Microstructures with a solid volume fraction within the range $0.25 \pm 0.001$ are then identified; see \cref{fig:latent_fish}. Among these, the microstructure with the highest value of the trace of permeability tensor, i.e., highest $\hat{{C}}_{00} + \hat{{C}}_{11}$, is selected \cite{liakopoulos1965darcy}. The chosen microstructure has the following shape parameters $\bm{M^*} = \{a = 0.7158, b = 0.3757, m = 0.6039, n_1 = 1.4787, n_2 = 0.4349, n_3 = 0.5857\}$. As one can observe in \cref{fig:latent_fish}, it exhibits a fish-like shape. This particular microstructure will be used in the next numerical experiment. 

\subsection{Bent-Pipe}
\label{sec: val}
\begin{figure}[H]
 	\begin{center}
	\includegraphics[width=110mm,scale=1.,trim={0 0 0 0} ]{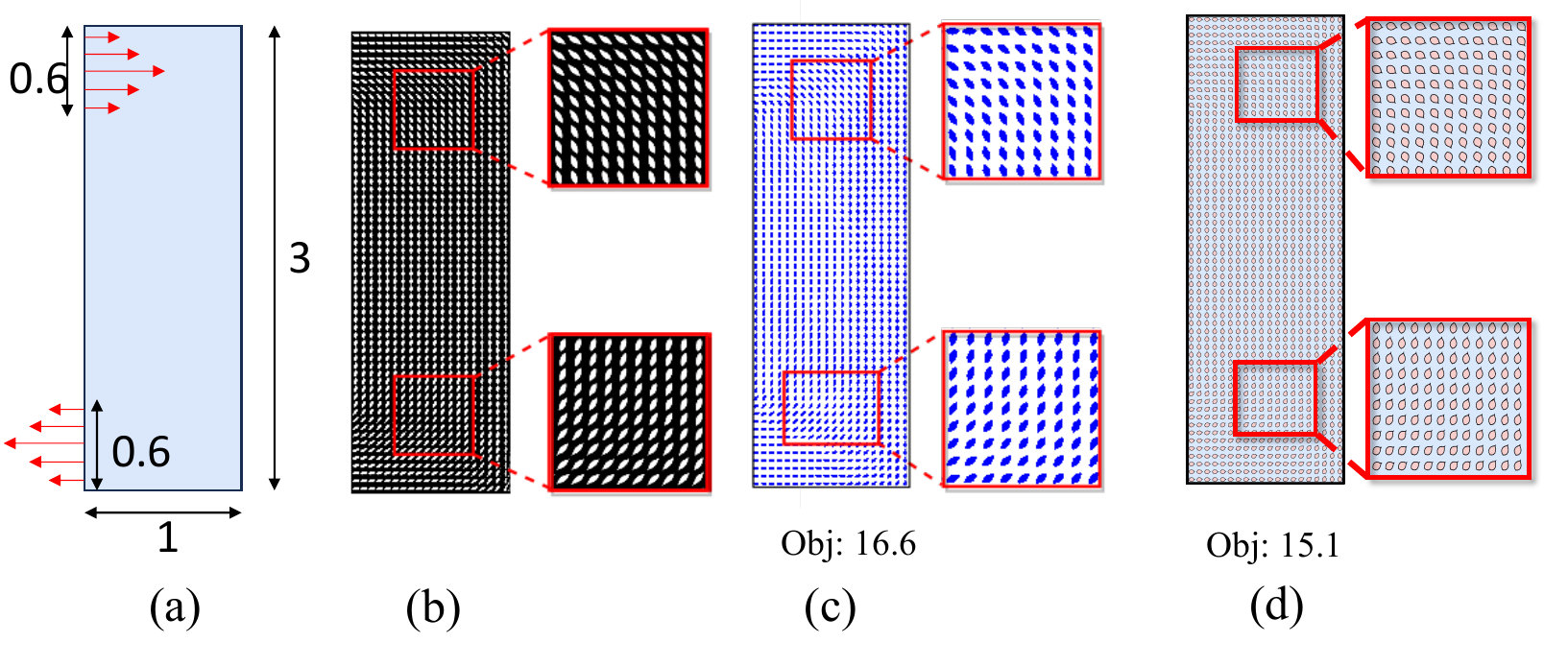}	
 \caption{Validation: (a) Problem definition. (b) Solution reported in \cite{wu2019topology}.  (c) Solution reported in \cite{padhy2023fluto}. (d) Topology generated via proposed method.}
	\label{fig:bent_val}
	\end{center}
 \end{figure}

We now consider the bent-pipe problem proposed in \cite{wu2019topology}, and illustrated in \cref{fig:bent_val}(a). The inlet and outlet boundaries are subject to parabolic velocity conditions, of unit magnitude, and the domain is discretized into $20 \times 60$ elements. In \cite{wu2019topology}, a two-scale topology optimization was carried out to minimize the dissipated power, with a constraint that the optimal microstructure must occupy exactly 25 percent of each unit cell. The reported topology is illustrated in \cref{fig:bent_val}(b); the final dissipated power was not reported. However, as noted in \cite{wu2019topology}, the computed microstructures resemble the fish-body. In \cite{padhy2023fluto}  a GMTO approach was employed with pre-defined microstructures to achieve a similar design as depicted in \cref{fig:bent_val}(c); the dissipated power was reported to be 16.6. Here we use the microstructure selected in the previous experiment to occupy each unit cell. Only the orientation of the microstructure in each cell is optimized. The resulting design is illustrated in  \cref{fig:bent_val}(d)  with the final dissipated power of 15.1. This experiment highlights that super-shapes sampled via the decoder can generate high-performing microstructural designs.

\subsection{Microstructure Variation}
\label{sec:mstr_var}

\begin{figure}[H]
 	\begin{center}
	\includegraphics[scale=0.5,trim={0 0 0 0} ]{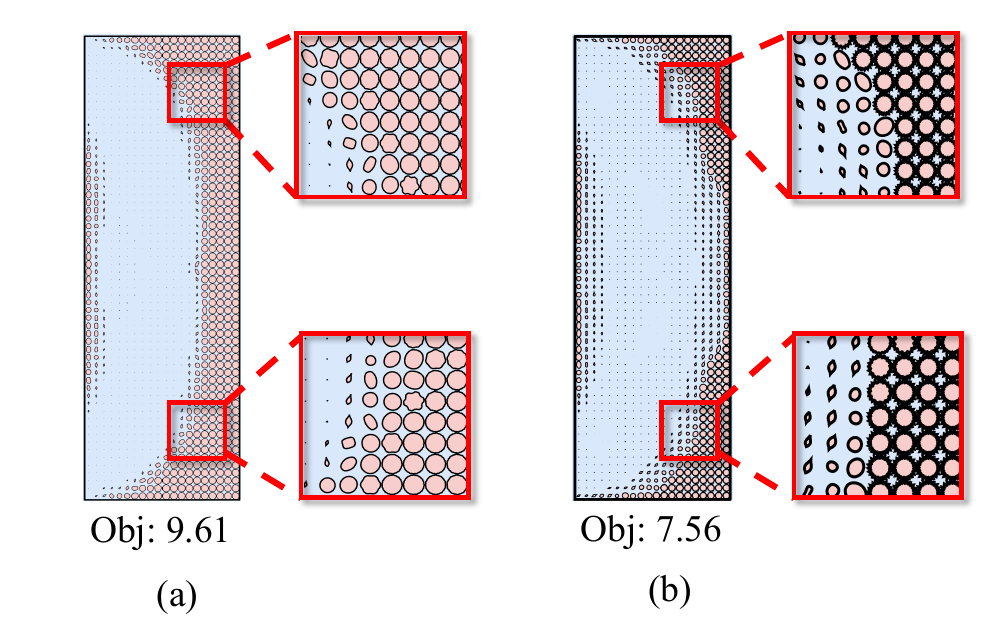}	
    \caption{Optimized design with microstructure variation with: (a) volume constraint, and (b) contact area constraint.}
	\label{fig:bent_mstr_var}
	\end{center}
 \end{figure}

We continue with the previous experiment, but we will now allow the shape and size of microstructures to vary across the domain. A global volume constraint of $0.75$ is imposed, as opposed to a unit-cell volume constraint.  The resulting design is illustrated in \cref{fig:bent_mstr_var}a with a dissipated power of $9.61$,  i.e., the performance improves with increased design space, as expected. The contact area for this particular design happens to be $75.69$.

Finally, instead of imposing a volume constraint, we impose a contact area constraint of $75.69$, and optimize the design. The final design is illustrated in \cref{fig:bent_mstr_var}b, with a dissipated power of $7.56$, i.e., further improvement in performance is achieved for the same contact area. 

Note that the dissipated power of $7.56$ and contact area of $75.69$ for the design in \cref{fig:bent_mstr_var}b are computed using the decoder. For validation, we re-computed the true values using a global FEA/homogenization of the final design. The dissipated power was found to be $7.87$ and the contact area was $78.49$, i.e., the decoder-reconstruction errors are relatively small.

\subsection{Convergence}
\begin{figure}[H]
 	\begin{center}
	\includegraphics[scale=0.3,trim={0 0 0 0} ]{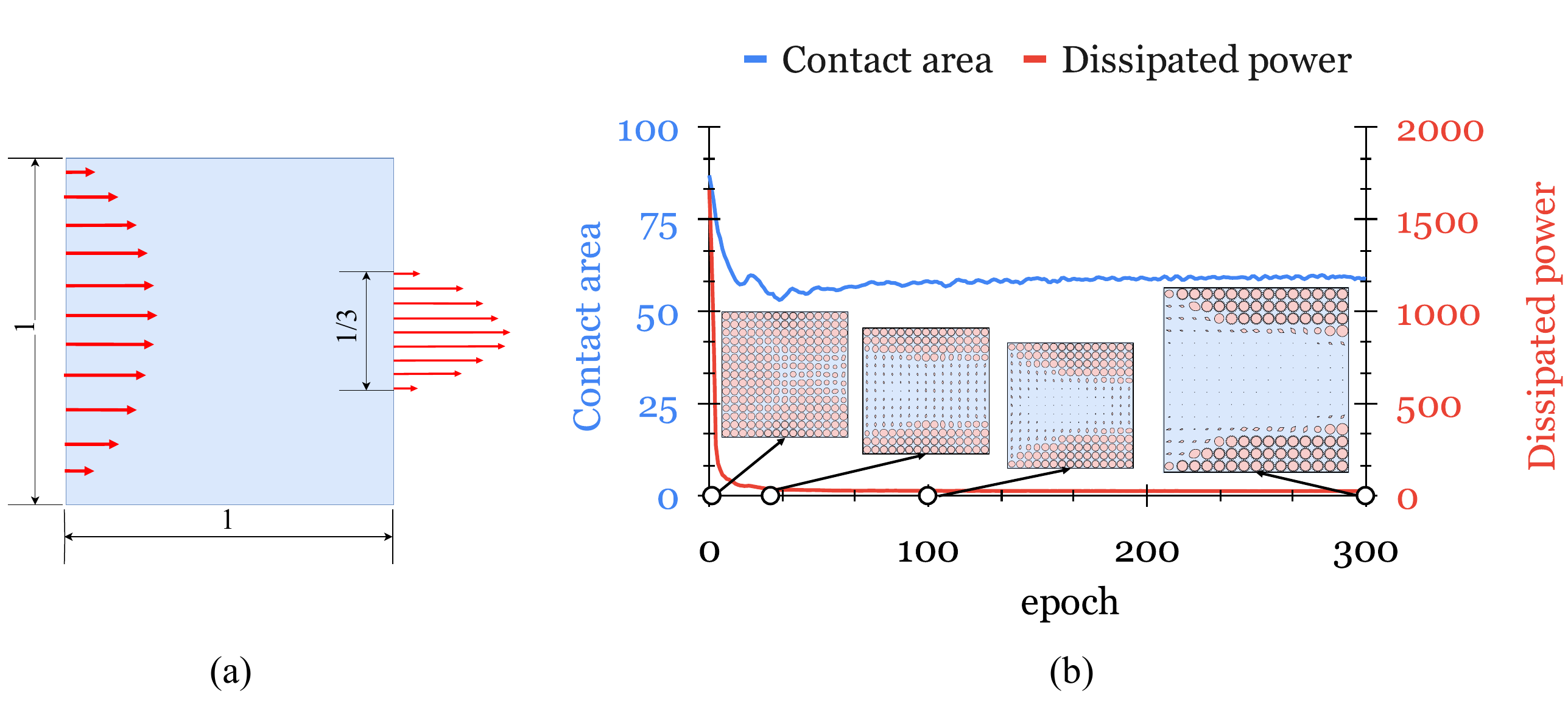}	
    \caption{Convergence of dissipated power and topologies for a diffuser problem. Topologies are illustrated at the 0th, 20th, 100th, and 300th (final) iterations. }
	\label{fig:convergence_diffuser}
	\end{center}
 \end{figure}

In this experiment, we demonstrate the typical convergence of the proposed algorithm using a diffuser problem, as shown in Figure  \cref{fig:convergence_diffuser}(a). The desired contact area was set to 60. The convergence of the dissipated power, contact area and the evolving topologies are illustrated in \cref{fig:convergence_diffuser}(b). We observed a stable convergence using the simple penalty formulation and Adam optimizer. Similar convergence behavior was observed for  other examples as well. 

 \subsection{Pareto trade-off}

\begin{figure}[H]
 	\begin{center}	\includegraphics[scale=0.25,trim={0 0 0 0} ]{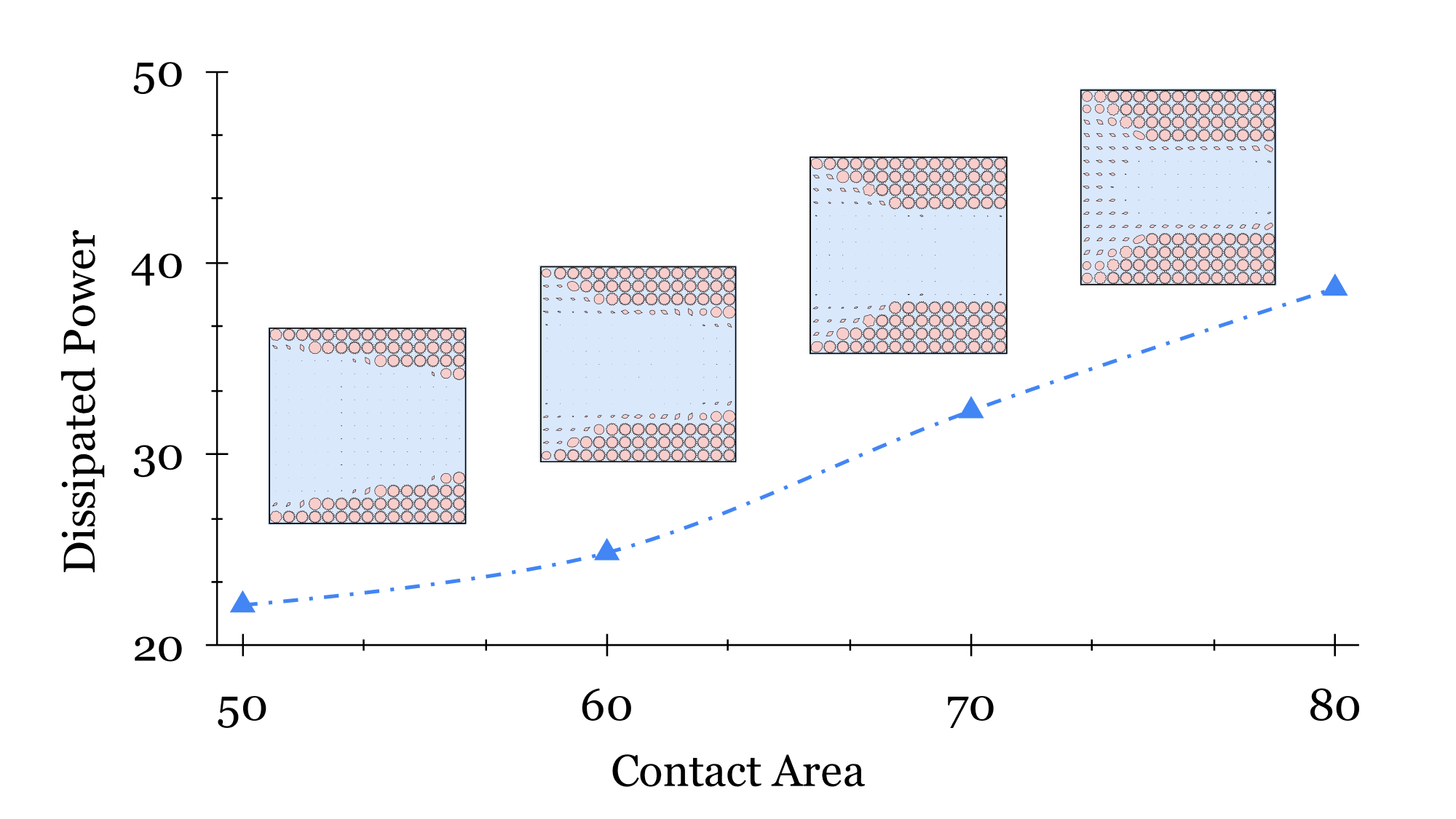}	
    \caption{Pareto }
	\label{fig:pareto}
	\end{center}
 \end{figure}

 Understanding the trade-off between the objective (dissipated power) and constraint (contact area) through exploration of the Pareto-front is crucial in making informed design choices. In this study, we considered the diffuser problem in \cref{fig:convergence_diffuser}(a), using the entire design space of microstructures. We computed the optimal topologies for different contact area constraints. \Cref{fig:pareto} illustrates that dissipated power increases with increasing contact area, as expected. 

 Once again, to determine the accuracy of decoder-reconstruction, we considered the design at the left-bottom corner in \cref{fig:pareto}. For this design, the decoder predicted value was $22.13$ for the dissipated power, and $50$ for the contact area. Using global FEA/homogenization, the  dissipated power was found to be  $22.72$ while the contact area was  $51.08$, i.e., the decoder-reconstruction errors of $2.6\%$ and $2.9\%$ are relatively small.

 \subsection{Computational cost}
One of the central hypotheses of this paper is that the proposed offline decoder-based framework offers significant computational advantage over concurrent homogenization based optimization. Here, we report the computational costs to validate this claim.

The offline homogenization and data-generation of 7000 microstructures (\cref{alg:DG}) required 164 minutes, while the VAE training   (\cref{alg:ED}) required  90 minutes, i.e., the total one-time overhead is around 250 minutes. The optimization of the bent-pipe (using a grid size of $20 \times 60$) took 32 minutes (300 iterations), while the optimization of the diffuser (using a grid size of $15 \times 15$) took 1.5 minutes (300 iterations).

We now consider a hypothetical scenario of  concurrent homogenization. From the above data, observe that the time required for each homogenization is $164/7000$, i.e., 1.4 seconds. Now for the concurrent homogenization of the bent-pipe, one must carry out homogenization over each cell within the grid  of $20 \times 60$ over 300 iterations, the expected optimization time is at least $1.4  \times 1200  \times 300/60 $, i.e., 8400 minutes. Similarly, for the diffuser, the expected optimization time is at least $1.4  \times 225 \times 300/60 $, i.e., 1575 minutes. Thus, the proposed offline decoder-based method is computationally far superior.

\subsection{Fabrication}
\label{sec:fab}
\begin{figure}[H]
 	\begin{center}	\includegraphics[scale=0.25,trim={0 0 0 0} ]{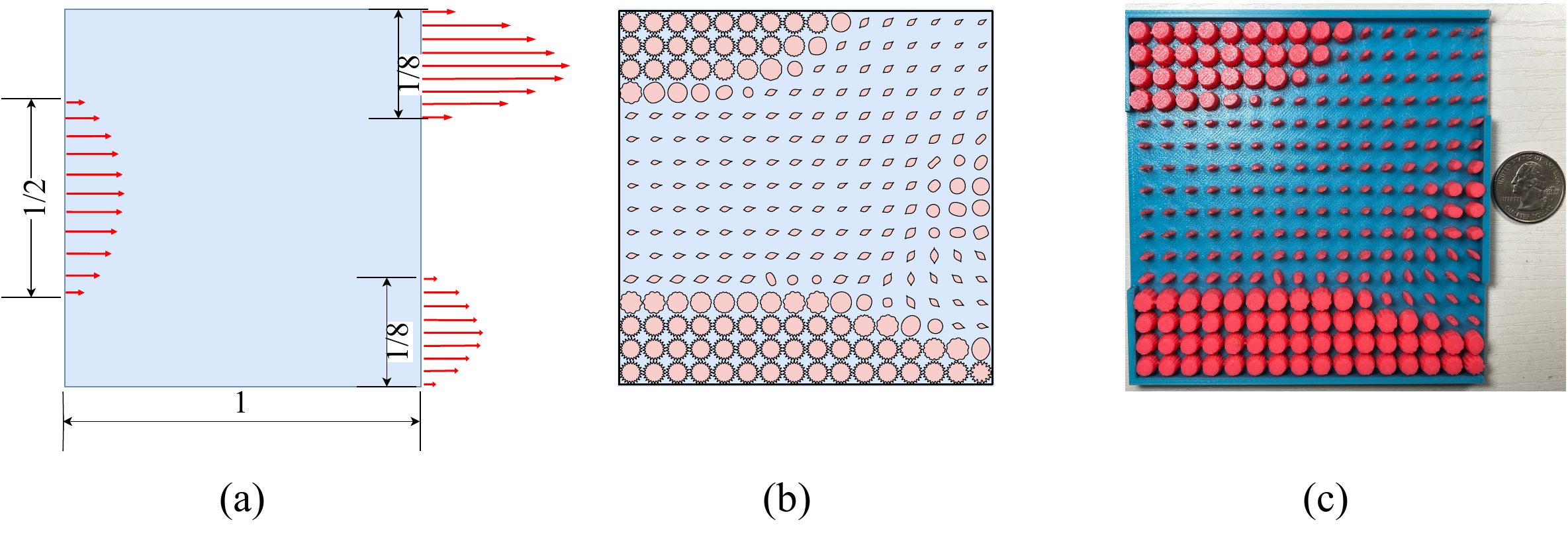}	
    \caption{(a) Design domain with boundary conditions, (b) optimized design, (c) 3d printed design }
	\label{fig:3d_print}
	\end{center}
 \end{figure}

To demonstrate the manufacturability of the designs produced by our framework, we consider a design domain with boundary conditions, as depicted in \cref{fig:3d_print} (a). In this example, we maintain a $1:1$ ratio and a $3:1$ ratio between the magnitude of the outlet velocity profiles and the inlet velocity profile. Additionally, we enforce a contact area constraint of 70, resulting in the design showcased in \cref{fig:3d_print} (b). To ensure that these designs can be manufactured successfully, we impose a minimum area constraint on each microstructure. The final 3D printed part is illustrated in \cref{fig:3d_print} (c).

\section{Conclusion}
\label{sec:conclusion}

In this paper, we presented a novel multi-scale fluid flow topology optimization framework using supershape  microstructures. An offline homogenization, along with the training of a VAE was used to generate a continuous and differentiable latent space of microstructural properties. This was followed by global optimization, where the dissipated power was minimized subject to contact area (or volume) constraint. 

The numerical results demonstrate that the proposed method is computationally far superior to concurrent homogenization, with minimal loss in accuracy. Furthermore, super-shapes increase the design space, yielding superior design compared to pre-defined microstructures.

Future research include extending the framework to (1) high Reynolds flow, (2) thermo-fluid applications, where the  contact area is determined indirectly via heat transfer, and (3) structural applications where microstructures with a genus greater than zero are desirable, and connectivity is also critical. Experimental validation, extension to 3D, and imposition of additional manufacturing constraints are also desirable. 

\section*{Acknowledgments}
This work was supported by the National Science Foundation grant CMMI 1561899.

\section*{Compliance with ethical standards}
The authors declare that they have no conflict of interest.

\section*{Replication of Results}
The Python code is available at \href{https://github.com/UW-ERSL/TOMAS}{github.com/UW-ERSL/TOMAS}

\bibliographystyle{unsrt}  
\bibliography{references}

\end{document}